\documentclass{PoS}

\title{Heavy ion theory: Review of recent developments with a first principles bias}

\ShortTitle{Heavy ion theory: Brief review}

\author{\speaker{Aleksi Vuorinen}\\
        Helsinki Institute of Physics and Department of Physics, P.O.~Box 64, FI-00014 University of Helsinki, Finland\\
        E-mail: \email{aleksi.vuorinen@helsinki.fi}}

\abstract{We discuss a number of prominent theoretical challenges in the physics of ultrarelativistic heavy ion collisions, and review some recent attempts to tackle them. These examples cover most stages of the collision process, but emphasis is given to approaches that rely on first principles methods or well-established Effective Field Theory frameworks. The topics discussed in most detail include the early dynamics of a collision as well as the properties of quark gluon plasma in thermal equilibrium.}

\FullConference{International Symposium on Lepton Photon Interactions at High Energies\\
		17-22 August 2015\\
		University of Ljubljana, Slovenia}

\begin{document}

\section{Challenges in heavy ion theory}

Gaining quantitative control over the physical processes taking place in an ultrarelativistic heavy ion collision undoubtedly ranks among the technically most challenging tasks within modern particle and nuclear physics. Reasons for its complexity are obvious: The system one is trying to describe is extensive, undergoes rapid and extremely nontrivial dynamics, and is fundamentally nonperturbative in nature. Subsequently, the current ``Standard Model of heavy ion physics'' \cite{Brambilla:2014jmp} is in effect a patchwork of effective descriptions of different kinds, each typically designed to describe a given epoch of the collision process and only applicable for a restricted class of physical observables. Notable examples of such effective theories (or models) include the Color Glass Condensate (CGC) framework for the initial state and early dynamics of the collision \cite{Gelis:2010nm}, the hydrodynamic description of the expansion of a locally equilibrated (or at least hydrodynamized) plasma \cite{Gale:2013da}, and the holographic mapping of strongly coupled dynamics to a gravitational system in five-dimensional Anti-de Sitter spacetime \cite{CasalderreySolana:2011us}. 

When preparing a compact review of a topic as extensive as theoretical heavy ion theory, it is almost mandatory to choose a bias of some kind unless one merely aims at constructing a list of references to recent works. In the case of the article at hand, this bias is chosen keeping in mind the gap that has traditionally existed between first principles calculations and heavy ion phenomenology. To this end, we have singled out a number of nontrivial physical problems encountered in different stages of the collision, where significant progress has recently been achieved using first principles approaches supplemented with robust Effective Field Theory tools. At times, this implies that we will find ourselves dealing with an unphysical limit or idealization of the system, but in each of the cases considered, making direct contact with phenomenology is no longer merely a distant goal for first principles calculations. 

Even with the first principles bias, it is clear that we will only be able to cover a small fraction of all interesting recent developments related to the topics discussed. We will thus refer the reader to the far more comprehensive review \cite{Brambilla:2014jmp} for a full list of appropriate references.

Our review is structured according to the different main branches of heavy ion physics, loosely ordered according to the evolution of a collision:
\begin{itemize}
 \item In section 2, we discuss the theoretical description of the early equilibration dynamics of a collision in the limits of asymptotically weak and strong coupling, commenting also on recent attempts to extend the corresponding approaches to intermediate couplings.
 \item In section 3, we review the current status of lattice and perturbative studies of the thermodynamic properties of quark gluon plasma (QGP) in equilibrium, covering both bulk and transport quantities.
 \item In section 4, we briefly discuss recent advances in the effective hydrodynamic description of a heavy ion collision.
 \item In section 5, we provide a few examples of hard probes of the QGP, discussing their theoretical description with first principles tools.
 \item In section 6, we conclude our presentation with our view of the current status of the field and the most promising and exciting future developments in it.
\end{itemize}

\section{Equilibration dynamics\label{sec:ED}}

A quantitative description of the initial state of a heavy ion collision --- let alone its evolution towards local thermal equilibrium --- is one of the most formidable challenges in heavy ion physics. According to common lore, first principles tools can only serve as qualitative aids in this process, as the system is so far from any of the limits where they are applicable, such as thermal equilibrium (relevant for lattice QCD) or asymptotically weak or strong coupling (perturbative QCD and holography, respectively). In recent years, this view has, however, been challenged, as thermalization studies at both strong and weak coupling have produced results in qualitative agreement with expectations based on comparison of experimental data with the results of hydrodynamic simulations. We will report on these developments below.

Due to confinement, the study of the initial state of a heavy ion collision is a fundamentally nonperturbative problem, and any (semi-)analytic approach to it is bound to involve a degree of model building. It turns out that for kinematic reasons, the most interesting part of the wave functions of the colliding nuclei --- responsible for a large fraction of the particles observed at mid-rapidity --- is that of small $x$. To describe this regime, one needs to be able to account for the physical effect of gluon saturation, i.e.~the overoccupation of the infrared (IR) gluonic degrees of freedom. The standard framework for the study of these modes is that of the Color Glass Condensate (CGC) \cite{McLerran:1993ni}, where a central observation is that the dynamics of the soft gluon fields becomes essentially classical at weak coupling, enabling the use of classical lattice simulations. In this limit, the framework becomes systematically improvable, and its study has indeed grown into a field of considerable magnitude (for a recent review, see e.g.~\cite{Lappi:2015jka}).

\begin{figure}[t]
\begin{center}
\includegraphics[width=8cm]{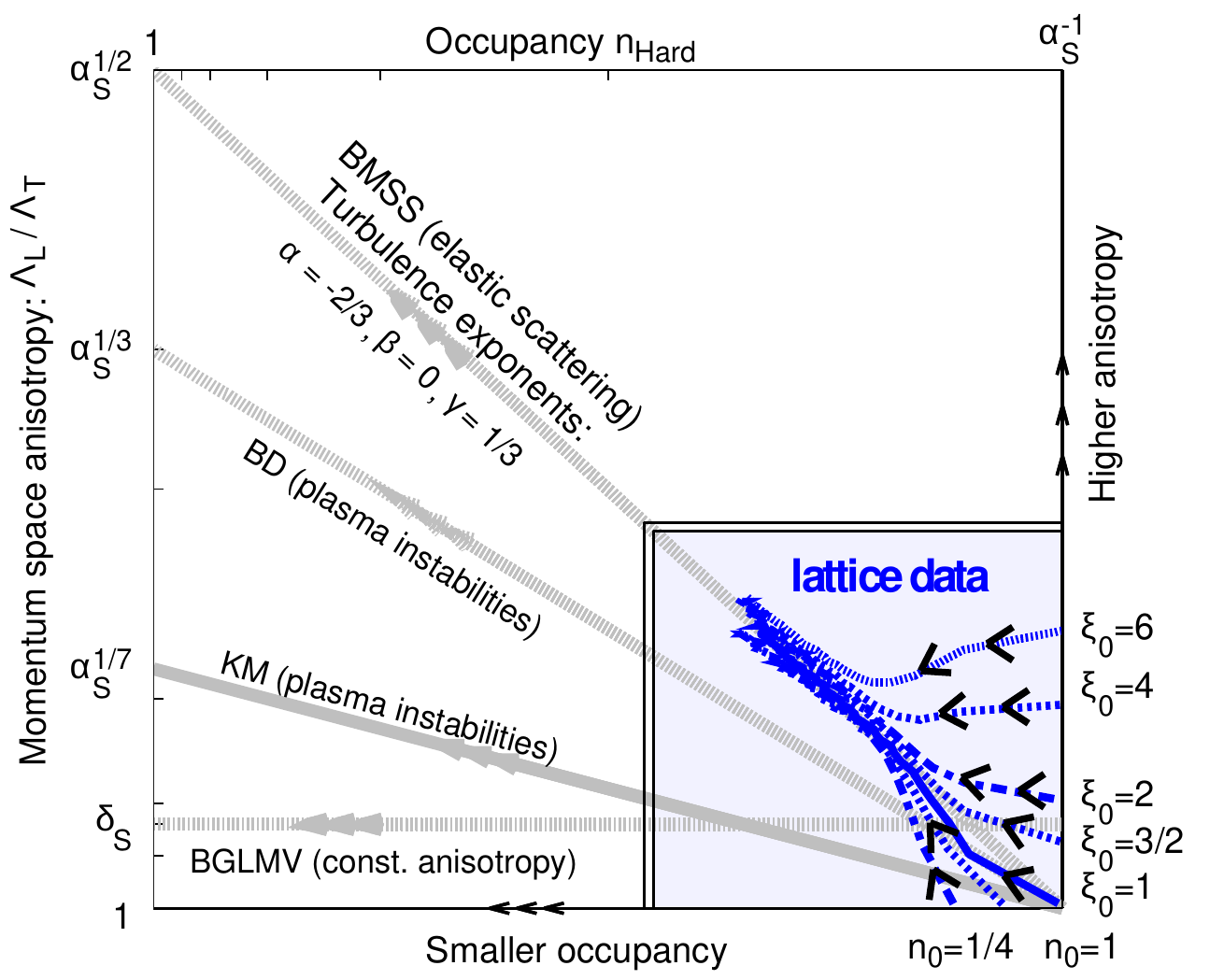}
\caption{In this plot, taken from  \cite{Berges:2013eia}, it is demonstrated that the classical evolution of a longitudinally expanding SU(2) Yang-Mills plasma is described by self-similar evoluation with scaling exponents consistent with the BMSS scenario \cite{Baier:2000sb}. For details of the notation, see \cite{Berges:2013eia}.}\label{fig:cartoon}
\end{center}
\end{figure}

One of the central questions in theoretical heavy ion physics is, how the system makes the transition from a CGC-type initial state into a near-thermal plasma that is amenable to a hydrodynamic treatment. According to standard lore, at asymptotically high energies --- or very weak coupling --- this process follows the so-called bottom-up scenario by Baier, Mueller, Schiff and Son (commonly referred to as the BMSS scenario), where soft gluons first form a thermal bath, followed by the radial break-up of the hard particles via elastic collisions \cite{Baier:2000sb}. The picture has since then been challenged by works emphasizing the role of plasma instabilities \cite{Kurkela:2011ti} or proposing the formation of a Bose-Einstein condensate in the system \cite{Blaizot:2011xf}, but classical lattice simulations appear to have vindicated the old BMSS scenario \cite{Berges:2013eia}. In fig.~\ref{fig:cartoon}, we demonstrate this result through a plot taken from \cite{Berges:2013eia}, where the time evolution of a longitudinally expanding classical SU(2) Yang-Mills system is displayed, starting from initial conditions with varying degrees of anisotropy. It is seen that after a brief initial regime, each of the solutions approaches a nonthermal fixed point, where the system exhibits self-similar behavior with scaling exponents predicted by the BMSS scenario. Whether this behavior could at later times still convert to e.g.~the one predicted by \cite{Kurkela:2011ti} is at the moment an open question.

The fact that the classical evolution of an initially anisotropic and out-of-equilibrium system leads to a nonthermal fixed point should by itself make it clear that something more --- namely, quantum effects --- are needed for the system to reach thermal equilibrium. The breakdown of the classical description can be understood from the fact that upon longitudinal expansion, the system eventually loses its overoccupied nature, and thus the gluon fields cease to be well-described by the solutions of the classical equations of motion. A crucial question then becomes, how one should describe the further evolution of the system in a way that takes into account all the necessary ingredients. At the moment, the two leading first principles tools available for this are Effective Kinetic Theory (EKT) as formulated by \cite{Arnold:2002zm}, as well as the gauge/gravity duality \cite{Maldacena:1997re} (see also \cite{CasalderreySolana:2011us,Gubser:2009md} for reviews). They become reliable in two opposing limits: EKT assumes that the system can be described in terms of weakly coupled quasiparticles, and requires the coupling constant of the theory to be small for the weak coupling expansion to converge. At the same time, holography is a tool derived for the description of an altogether different theory than QCD, namely ${\mathcal N}=4$ Super Yang-Mills (SYM), and even there practical calculations typically assume the limits of large 't Hooft coupling $\lambda\equiv g^2 N_c$ and number of colors $N_c$. Despite these limitations, a considerable amount of qualitative insight has been drawn from both of these types of calculations, as we will presently review.

Most of the applications of EKT to a thermalizing gauge theory plasma are very recent, owing to the fact that for a long time it was believed that the leading order bottom-up description of the thermalization of a weakly coupled system is orders of magnitude too slow to be of phenomenological relevance for heavy ion collisions. This view was, however, recently challenged in a paper \cite{Kurkela:2014tea}, which demonstrated that when extrapolated to phenomenologically relevant couplings, a full leading order calculation of the weak coupling thermalization time gives estimates of phenomenologically relevant magnitude, of order 1 fm/c. Very recently, a numerical EKT study of a longitudinally expanding system has furthermore demonstrated that at $\lambda=10$ the plasma becomes amenable to a hydrodynamical description after 1-2 fm/c \cite{Kurkela:2015qoa}. This result is, however, rather strongly dependent on the value of the coupling, and the main uncertainty in calculations of this kind indeed originates from their sensitivity to higher order corrections in the formulation of the EKT framework.

\begin{figure}[t]
\begin{center}
\includegraphics[width=10cm]{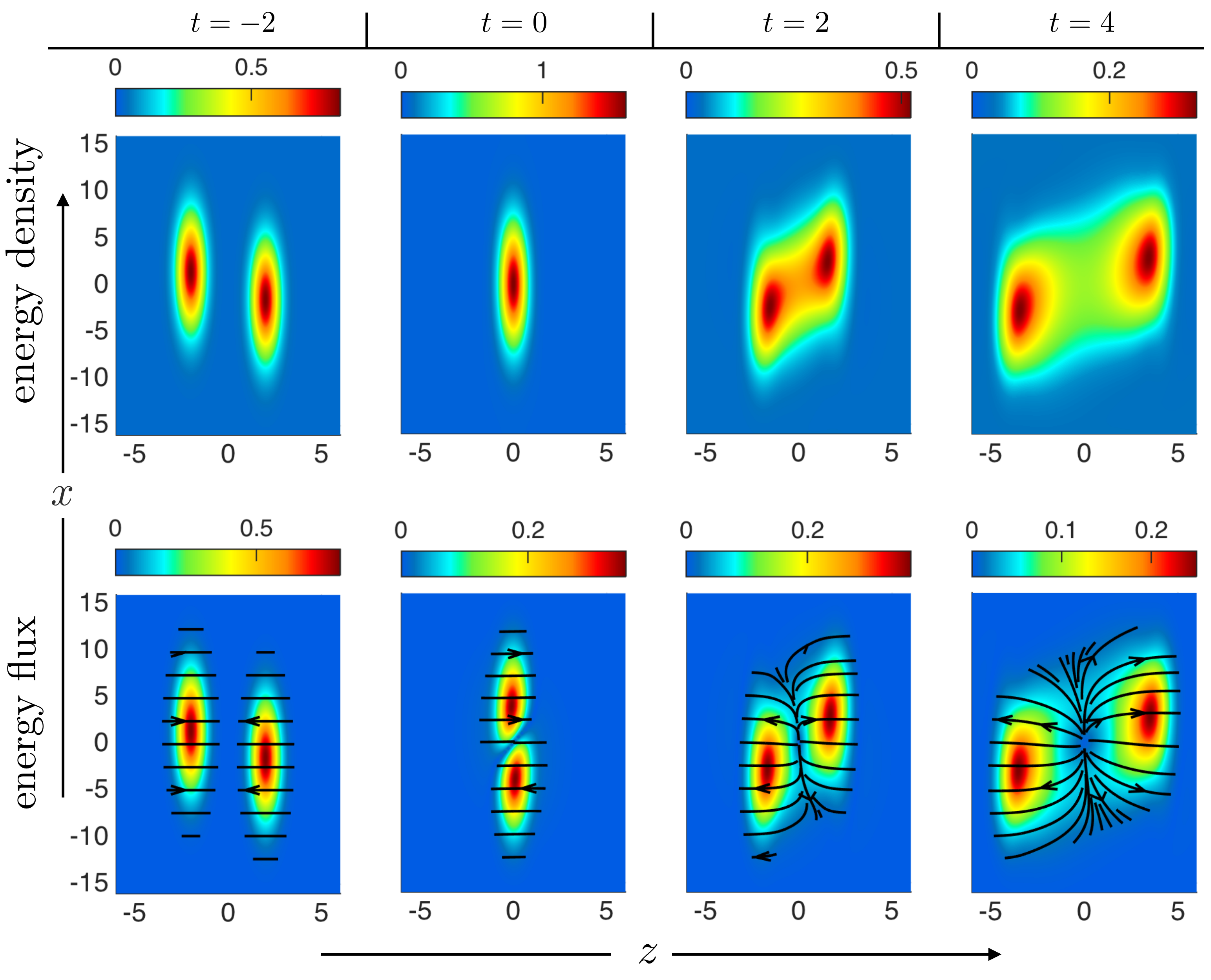}
\caption{In this plot, taken from \cite{Chesler:2015wra}, we display four snapshots of the time evolution of a collision of two localized shock waves in strongly coupled ${\mathcal N}=4$ SYM theory, giving the behaviors of the energy density and energy flux. The emergence of transverse flow is clearly visible in the figure.}\label{fig:snapshots}
\end{center}
\end{figure}

In the opposite limit of an infinitely strongly coupled SYM theory, holography offers a dual description of the thermalization process in terms of gravitational collapse and black hole formation in asymptotically Anti-de Sitter (AdS) spacetimes. Subsequently, a lot of work has been invested into the development of tools for the study of gravitational dynamics in such spacetimes, resulting in detailed studies of topics such as isotropization dynamics \cite{Chesler:2008hg,Heller:2012km,Heller:2013oxa,Fuini:2015hba} and collisions of planar shock waves \cite{Chesler:2010bi,Casalderrey-Solana:2013aba,Casalderrey-Solana:2013sxa,vanderSchee:2013pia,Chesler:2015fpa}. Recently, these calculations have progressed to a point, where it is even possible to consider the collisions of two fully localized shock waves with nonzero impact parameter \cite{Bantilan:2014sra,Chesler:2015wra,Chesler:2015bba}, of which we show an example in fig.~\ref{fig:snapshots}. In addition, significant progress has been achieved in the determination of off-equilibrium Green's functions in these backgrounds \cite{CaronHuot:2011dr,Keranen:2014lna}, resulting in studies of physical effects such as particle production in a thermalizing plasma \cite{Baier:2012tc,Baier:2012ax,Baron:2012fv,Lin:2013sga}.

So far, the most important lessons that have been drawn from studies of holographic thermalization are that strong coupling thermalization occurs naturally at a time scale of $1/T$ --- leading to phenomenologically relevant values in heavy ion collisions --- and that it proceeds in the top-down fashion. Another very interesting observation is the early onset of hydrodynamic behavior: Hydrodynamization appears to occur in the system much before isotropization, which was not expected before the first simulations of shock wave collisions were carried out \cite{Chesler:2010bi}. Recent studies of black hole formation have in addition stressed the role of universality of black hole formation in attempts to draw lessons about strong coupling thermalization.

Challenges facing the holographic description of thermalization include most importantly relaxing the limits of large $\lambda$ and $N_c$ as well as conformal invariance. Of these, the most pertinent problem appears to be the approximation related to the magnitude of the 't Hooft coupling, as it is by far not obvious that at phenomenologically relevant values of $\lambda\sim 20$, corrections originating from an expansion in its inverse powers would be negligible. Indeed, the few thermalization calculations that have so far been performed including finite coupling corrections \cite{Baron:2012fv,Steineder:2012si,Steineder:2013ana} have given indications of sizable corrections in the relevant region of 't Hooft couplings. Recently it has, however, been demonstrated that at least for quasinormal mode frequencies, responsible for the late time behavior of a thermalizing system, a simple resummation of higher order corrections is enough to significantly improve the convergence of strong coupling expansions \cite{Waeber:2015oka}.

Although a lot has been recently achieved in the study of the early thermalization dynamics of heavy ion collisions, much remains to be done. On the holographic side, in addition to finite coupling effects, attention should be paid on the effects of conformality breaking, as pioneered by \cite{Craps:2013iaa,Janik:2015waa,Gursoy:2015nza}. Alongside with these developments, it would be highly interesting to see, how the weak coupling thermalization picture gets altered by the inclusion of Next-to-Leading Order (NLO) corrections to the EKT setup. Finally, a quantitative comparison of the characteristics of weakly and strongly coupled thermalizing systems, cf.~e.g.~\cite{Kurkela:2015qoa} and \cite{Chesler:2015wra}, might reveal interesting lessons about universality in equilibration dynamics.

\section{Bulk thermodynamics of the quark gluon plasma\label{sec:bulk}}

From a purely phenomenological point of view, the most interesting aspects of the equilibrium thermodynamics of the QGP have to do with the fact that hydrodynamical simulations require information about the Equation of State (EoS) and viscosities of the plasma as input. At the same, it should, however, be kept in mind that many of the most fundamental questions that the entire heavy ion physics program is trying to address --- such as the structure of the QCD phase diagram and the location of a possible tricritical point therein --- are related to equilibrium physics. Thus, it is clearly of high interest to investigate, how these questions are addressed on the theoretical side using methods such as lattice QCD and weak coupling expansions.

\begin{figure}[t]
\begin{center}
\includegraphics[width=6.5cm]{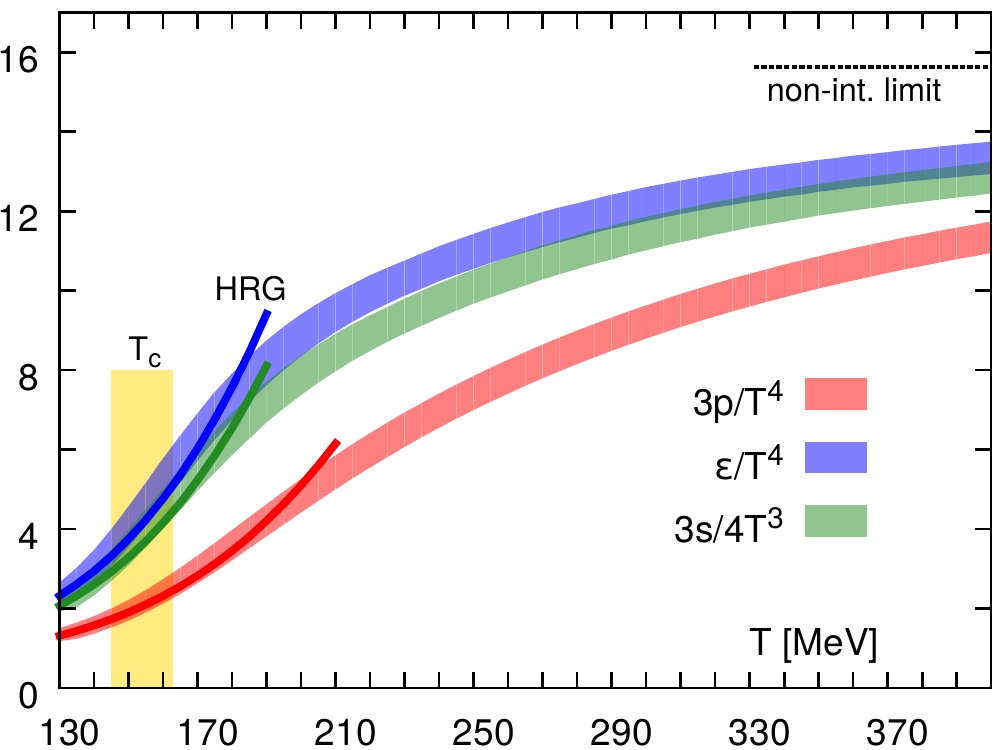}$\;\;\;\;\;\;\;\;\;$\includegraphics[width=6.5cm]{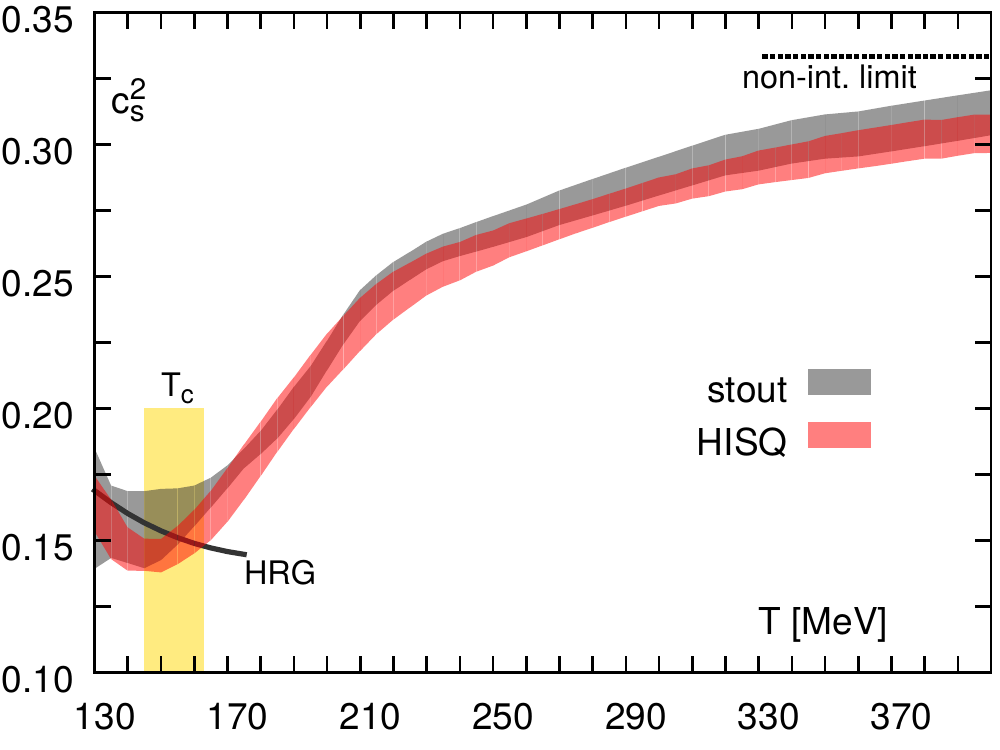}
\caption{The behavior of the continuum-extrapolated pressure, energy density and entropy (left figure) as well as the speed of sound squared (right figure) in 2+1 flavor QCD, taken from \cite{Bazavov:2014pvz}. Quantitatively similar results have been reported also in \cite{Borsanyi:2013bia}.}\label{fig:hoteos}
\end{center}
\end{figure}

As the only nonperturbative first principles tool available for QCD, lattice Monte-Carlo simulations are clearly the method of choice in the study of QGP thermodynamics as long as they are feasible. In the context of equilibrium physics, their main restriction originates from finite baryon density due to the famous Sign Problem \cite{Petreczky:2013qj}. At vanishing quark chemical potentials $\mu_f$, both the EoS and various quark number susceptibilities, as well as the details of the deconfinement and chiral phase transitions, can however be reliably determined. They have indeed been addressed by several lattice groups, which by now have reached a consensus over most of the relevant questions. The chiral and deconfinement transitions are found to be both of crossover nature and to occur at temperatures around 150-160 MeV \cite{Aoki:2009sc,Bhattacharya:2014ara}. The behavior of the pressure, energy density, entropy and speed of sound in 2+1 flavor QCD are on the other hand depicted in fig.~\ref{fig:hoteos}, where a particularly interesting aspect is the good agreement of the lattice results with those of a hadron resonance gas calculation at temperatures around $T_c$. Similarly impressive agreement between the lattice data and the predictions of resummed perturbation theory is observed at temperatures above ca.~3$T_c$ (cf.~\cite{Kajantie:2002wa,Andersen:2011sf} for the state-of-the-art perturbative calculations).

\begin{figure}[t]
\begin{center}
\includegraphics[width=7.2cm]{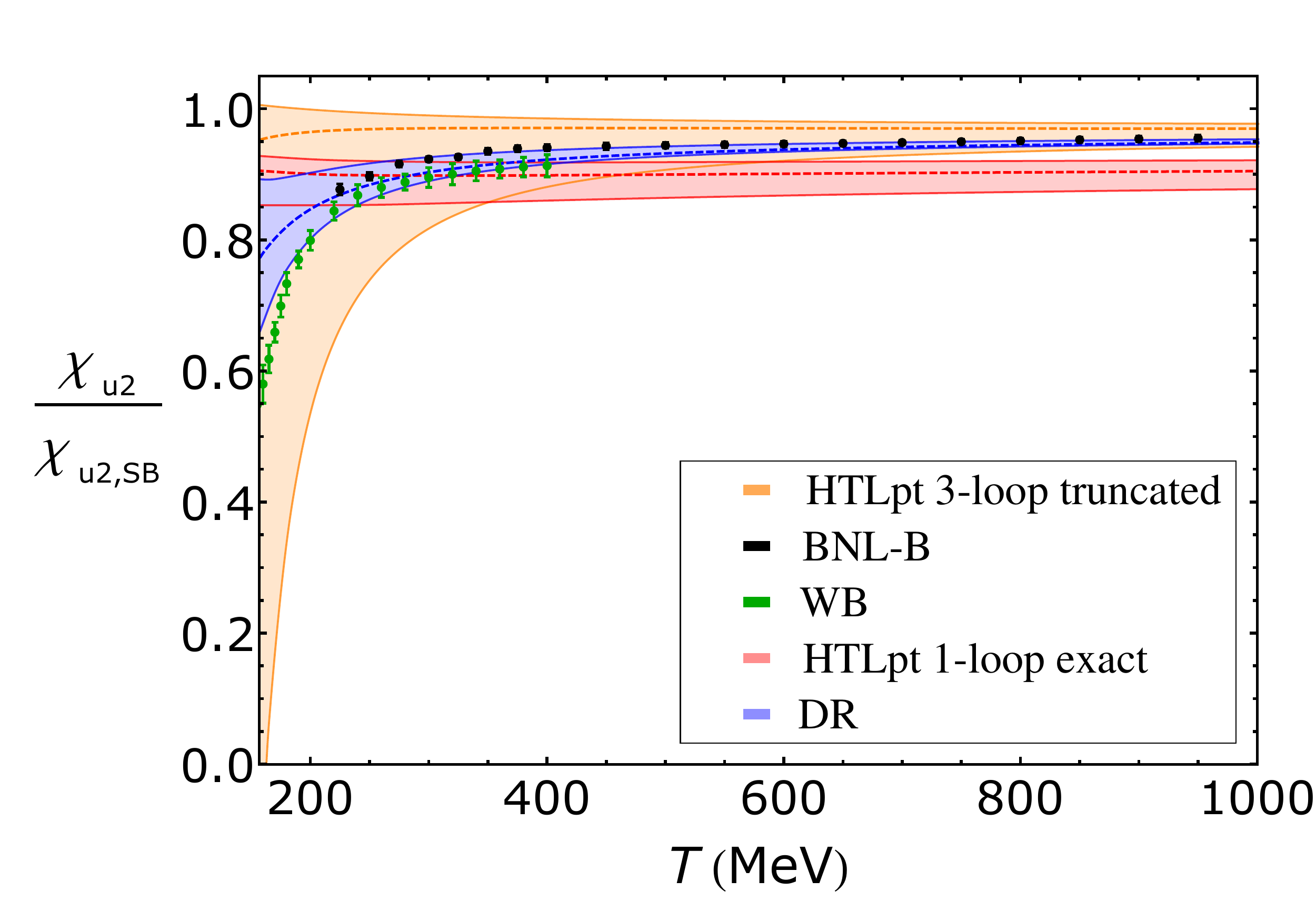}\includegraphics[width=7.5cm]{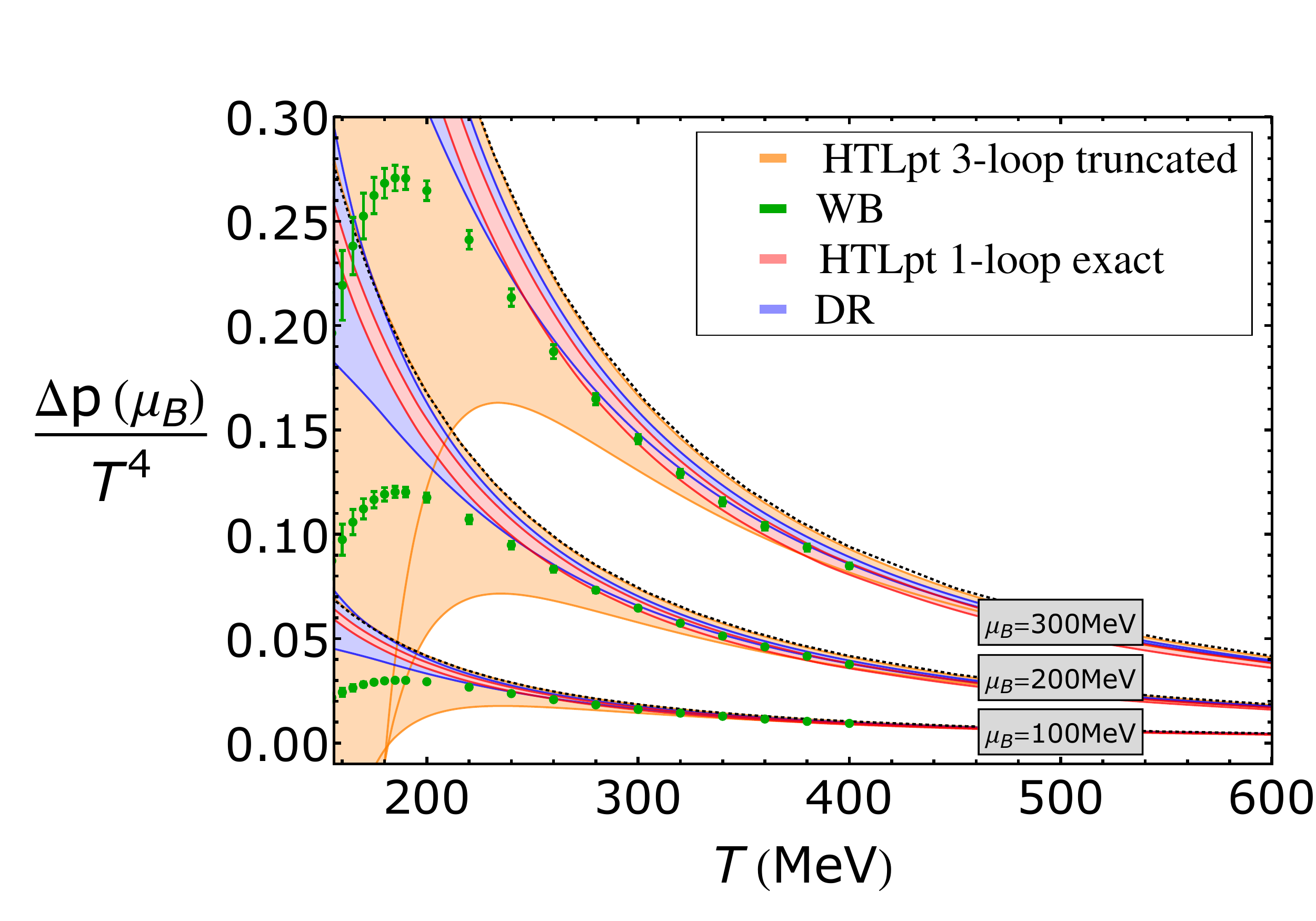}
\caption{The second order diagonal quark number susceptibility of a light quark (left figure) and the density dependent part of the QCD pressure (right figure), both given as functions of temperature. The figures are from \cite{Mogliacci:2013mca}, while the lattice data shown has been taken from \cite{Bazavov:2013uja,Borsanyi:2011sw}.}\label{fig:suscs}
\end{center}
\end{figure}

Next, we move on to quantities that probe the effects of finite density, such as quark number susceptibilities or the pressure at small chemical potentials (obtained on the lattice by Taylor expanding the quantity around $\mu_f=0$). These have been determined on the lattice e.g.~ in \cite{Karsch:2010hm,Borsanyi:2012cr}, the results of which we compare to those of resummed perturbation theory \cite{Vuorinen:2003fs,Haque:2014rua,Mogliacci:2013mca} in fig.~\ref{fig:suscs}. Again, we observe excellent agreement from rather low temperatures onwards for both the 3-loop Hard Thermal Loop perturbation theory (HTLpt) result and another one applying a Dimensional Reduction \cite{Kajantie:1995dw} inspired resummation scheme to the partial 4-loop results of ordinary perturbation theory (cf.~\cite{Laine:2006cp} for the introduction of the method). The resummations clearly play a key role in this success of weak coupling expansions, as the predictions of unresummed perturbation theory come with such a high degree of sensitivity to the renormalization scale that it is only at extremely high energy densities that any resonable predictions can be made.

With only the LHC heavy ion collisions in mind, there would be little phenomenological motivation to inspect densities higher than what can be studied on the lattice with the Taylor expansion method. The situation is, however, completely different, if one is interested in mapping the phase diagram and studying the possible existence of a tricritical point, relevant for the future heavy ion experiments planned at FAIR. Here, one is unfortunately faced with a lack of first principles options, as lattice QCD is restricted to small densities by the convergence of the Taylor expansion, and (even resummed) perturbation theory is on the other hand incapable of accessing energy densities close to the critical region. A lot of attention has thus been placed on studying a variety of models that are hoped to capture the essential physics in this region (cf.~e.g.~\cite{Fukushima:2008wg,Fukushima:2010bq} for topical reviews), while numerous attempts have been made to cure the Sign Problem of lattice QCD (see e.g.~\cite{Aarts:2008wh,Cristoforetti:2012su} and references therein).

Apart from high density, another highly problematic area for lattice QCD is the determination of genuinely Minkowskian quantities, such as different spectral functions and the associated transport coefficients. A quantity of particular relevance for heavy ion phenomenology is clearly the shear viscosity of the QGP, whose introduction to hydrodynamic codes has turned out to be mandatory \cite{Romatschke:2007mq}. Despite extensive efforts, a direct lattice determination of this quantity has only been possible with considerable systematic uncertainties \cite{Meyer:2007ic} (see also \cite{Meyer:2011gj}), and to this end alternative methods have been widely investigated. One promising recent idea is to use perturbative information about the ultraviolet behavior of the spectral functions to aid the analytic continuation of lattice data for the Euclidean imaginary time correlator \cite{Burnier:2011jq}. This method has been successfully applied to the estimation of the flavour diffusion coefficient of the QGP \cite{Burnier:2012ts}, while efforts to generalize these results to other transport quantities, such as the bulk and shear viscosities, are currently underway (cf.~\cite{Laine:2011xm,Zhu:2012be,Vuorinen:2015wla} for related perturbative work and discussion).

\begin{figure}[t]
\begin{center}
\includegraphics[width=10.0cm,height=7.0cm]{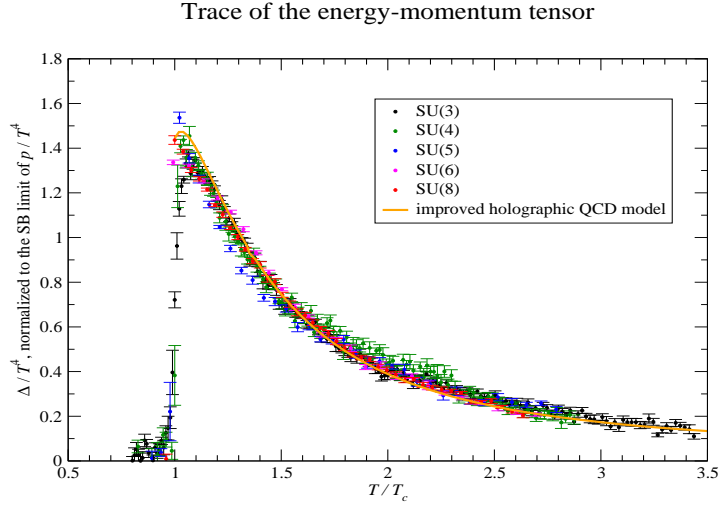}
\caption{The behavior of the trace anomaly of SU($N_c$) Yang-Mills theory as computed on the lattice \cite{Panero:2009tv} and in the IHQCD model \cite{Gursoy:2007cb}. The figure is taken from \cite{Panero:2009tv}.}\label{fig:marco}
\end{center}
\end{figure}

Finally, we note that similarly to the the case of the early thermalization dynamics of a heavy ion collision, also the equilibrium properties of the QGP can be approached using holography. Here, a particularly popular approach has been to build phenomenological bottom-up models that by construction feature the running QCD gauge coupling, such as Improved Holographic QCD (IHQCD) \cite{Gursoy:2007cb}. These models have been shown to accurately reproduce the equilibrium thermodynamics of the theory (see e.g.~\cite{Kajantie:2006hv}), as verified in fig.~\ref{fig:marco}, where we display the comparison of the IHQCD prediction for the trace anomaly of large-$N_c$ Yang-Mills theory with properly normalized lattice results for various $N_c$ \cite{Panero:2009tv}. A generic prediction of these (so-called two-derivative) models is that the shear viscosity over entropy ratio equals $1/(4\pi)$, which indeed appears to be a good approximation close to $T_c$, but less so at higher temperatures.

\section{Hydrodynamic expansion of the plasma\label{sec:hydro}}

Equipped with information about the equilibrium properties of the QGP and with at least some qualitative understanding of the early dynamics of a heavy ion collision, it is natural to apply these insights into a hydrodynamic simulation of the longitudinal and transverse expansion of the fireball. In this context, it is important to recall that hydrodynamics is an effective description based on a derivative expansion that can be expected to converge when deviations from local thermodynamic equilibrium are small, i.e.~when the initially strongly nonlinear evolution of the system has had time to ``hydrodynamize'' it. At leading order, hydrodynamics describes the flow of an ideal fluid, but dissipation effects enters through shear and bulk viscosities at the NLO level. The most important qualitative effect of the hydrodynamic flow is that it converts the initial spatial anisotropy of the system into a momentum space one, measured in experiments as the so-called flow coefficients.

\begin{figure}[t]
\begin{center}
\includegraphics[width=7.2cm]{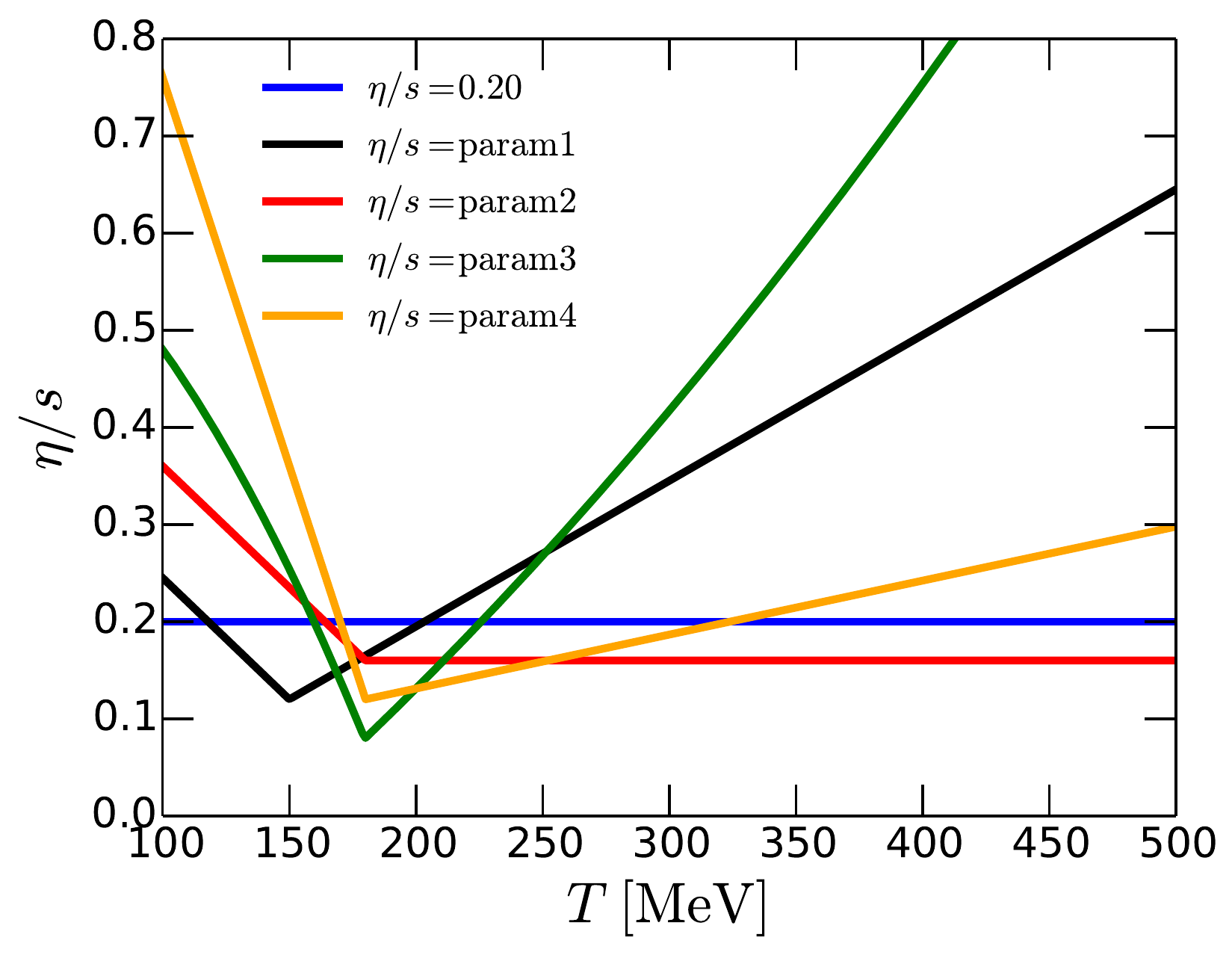}$\;\;\;\;$ \includegraphics[width=6.8cm]{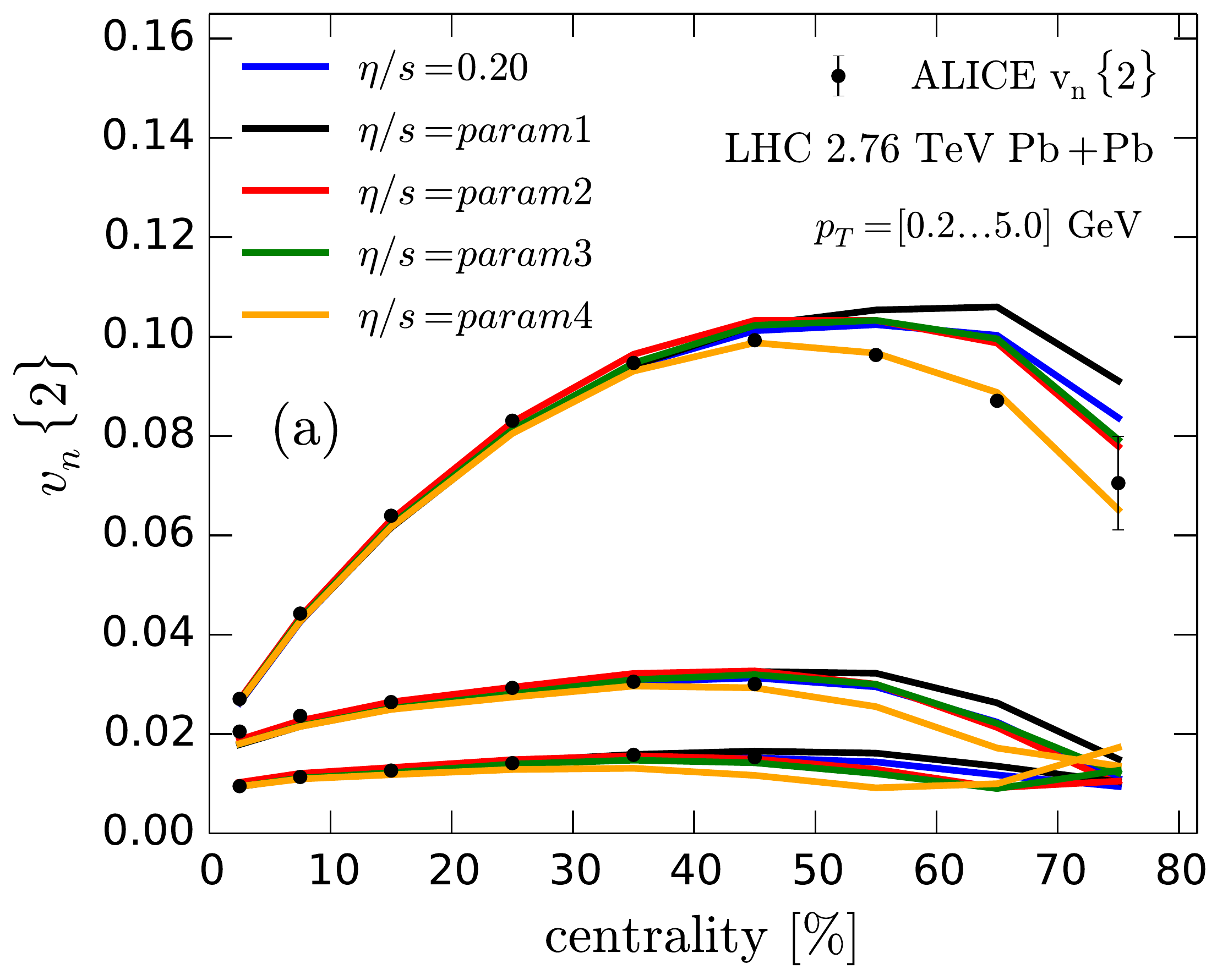}
\caption{Left: Five different parameterizations of the temperature dependence of $\eta/s$ used in the work of \cite{Eskola:2015uda}. Right: The centrality dependence of the flow coefficients $v_2$, $v_3$ and $v_4$ as determined in viscous hydrodynamic simulations, using the temperature-dependent values of $\eta/s$ indicated in the left plot. The figures are from \cite{Eskola:2015uda}, and the experimental data from \cite{ALICE:2011ab}.}\label{fig:hydro}
\end{center}
\end{figure}

During the past two decades, the line of work described above has grown into an industry of considerable magnitude, and in fact plays a crucial role in connecting first principles theoretical calculations with experimental data. Conversely, varying plasma parameters and initial conditions in the hydrodynamic simulations has provided important phenomenological information about the properties of the QGP, indicating apparently very fast thermalization dynamics as well as a remarkably low shear viscosity to entropy ratio \cite{Romatschke:2007mq}. These observations have indeed been the primary factors leading to the current consensus about the strongly coupled nature of the matter produced in HICs \cite{CasalderreySolana:2011us}. Below, we will provide a brief account of a few selected recent developments in hydrodynamical simulations, while for a more comprehensive review of the present status of the field, we refer the reader to \cite{Gale:2013da}.

One important development with hydrodynamical simulations is that they have recently become accurate enough to be sensitive to the temperature dependence of the shear viscosity as well as details of the EoS of the plasma. While the EoS is rather well predicted by lattice QCD (cf.~section 3), for the shear viscosity this opens up an intriguing possibility to determine its functional form through a comparison of hydrodynamic predictions with experimental data. This approach has indeed been followed in \cite{Eskola:2015uda}, where hydrodynamic simulations were carried out with five different parameterizations of the temperature dependence of $\eta/s$, cf.~fig.~\ref{fig:hydro}. As can be seen from the corresponding analysis of the centrality dependence of three different flow coefficients, a rather strong temperature dependence, consistent with a fast decrease of $\eta/s$ around $T_c$ and a slow increase thereafter, provides the best fit to data from the ALICE experiment. A similar comparison with RHIC data however favors constant behavior, so at the moment one should remain cautious about the conclusions to be drawn. Nevertheless, the study performed in \cite{Eskola:2015uda} --- as well as a similar analysis of the EoS in \cite{Pratt:2015zsa} --- demonstrates that in the future, it may well be possible to directly pinpoint many properties of the equilibrium QGP using hydrodynamic simulations.

Further recent developments in relativistic hydrodynamics include e.g.~the incorporation of a nonzero bulk viscosity in the simulations \cite{Ryu:2015vwa}, an extensive study of study of the effects of anisotropies in the hydrodynamical expansion \cite{Martinez:2012tu}, as well as the discovery of new exact solutions to the Boltzmann equation \cite{Denicol:2014tha}. These issues are discussed at length e.g.~in \cite{Gale:2013da}.

\section{Hard probes\label{sec:HP}}

All of the physical quantities we have discussed so far --- the EoS, viscosities, and even the thermalization time --- share the same shortcoming: The only way to connect them to experimental data is through hydrodynamic simulations, which makes them in effect \textit{soft} probes of the collision. There is, however, an altogether different class of observables that probes the plasma at considerably higher momentum scales, including jets, heavy flavor objects, as well as photons and (di)leptons emitted at high energies. These quantities are commonly dubbed hard probes of the plasma, and their theoretical study typically combines perturbative QCD --- applicable at the high momentum scales associated with the probes themselves --- with nonperturbative elements necessary for the description of the surrounding medium and (in the case of jets) the interactions of the probe with the medium. The special property of the hard probes is that alongside accessing high energies, they provide information about the pre-thermalization era of the collision, as the information carried away in particular by photons and leptons escapes the plasma almost freely. A recent review of the theoretical aspects of hard probes in heavy ion physics can be found in \cite{Schenke:2015nva}. 

In jet physics, a phenomenon specific to heavy ion collisions is that of jet quenching: When a highly energetic pair of back-to-back partons is formed near an edge of the collision region, it may happen that one of the partons forms a jet that escapes the plasma nearly unaltered, while its counterpart flying to the opposite direction must traverse a much longer time through the medium, thereby losing its energy. This process has a very specific experimental signature in terms of single jets, and its quantitative theoretical description is one of the main challenges in the study of hard probes. Reaching this goal however necessitates understanding energy loss in a moderately strongly coupled medium, which is a formidable task; in fact, so far a robust understanding of energy loss mechanisms has only been gained at extremely weak couplings (to leading order in perturbation theory) as well as in the holographic limit.

An important result that has facilitated considerable progress in the field of jet physics is the parameterization of the momentum broadening and collisional energy loss of partons (jets) in terms of one parameter $\hat{q}$, which is commonly dubbed the jet quenching parameter \cite{Baier:1996sk}. Subsequently, a large number of recent works have focused on attempts to determine its value using weak coupling techniques \cite{CaronHuot:2008ni,Laine:2012ht,Blaizot:2014bha}, a combination of lattice simulations and dimensionally reduced effective theory \cite{Panero:2013pla}, as well as the gauge/gravity duality \cite{Liu:2006ug}. Typical estimates for the quantity, of relevance for RHIC and LHC collisions, range between 5 and 10 GeV$^2$/fm, demonstrating the currently still sizable uncertainties in these calculations. 

Another example of important recent progress in the theory of hard probes originates from studies of thermal photon and dilepton emission in a QGP, which have been determined to NLO in the gauge coupling in several heroic computations \cite{Ghiglieri:2013gia,Laine:2013vma,Ghisoiu:2014mha,Ghiglieri:2015nba}. An interesting aspect of these calculations is that as long as we only have a few examples of dynamic quantities, for which a full NLO perturbative result exists, it is difficult to assess the convergence properties of the expansions. When extrapolated to couplings relevant for heavy ion collisions at the LHC, the results of \cite{Ghiglieri:2013gia} for photon production show that the NLO correction only amounts to a 20\% shift to the LO result, while in the case of the heavy quark diffusion coefficient --- the first ever dynamic quantity computed to NLO \cite{CaronHuot:2007gq} --- the corrections are rather at the 100\% level. 

Finally, it should be noted that photon and dilepton emission can also be studied using holography, where they have been initially computed in infinitely strongly coupled ${\mathcal N}=4$ SYM theory \cite{CaronHuot:2006te}, and later generalized to finite 't Hooft couplings \cite{Hassanain:2011ce} and even to an out-of-equilibrium setting \cite{Baier:2012tc,Baier:2012ax}. It is noteworthy that at least in the case of the ${\mathcal N}=4$ SYM theory, the transition from weak to strong coupling appears to be smooth, with the sharp peak of the perturbative photon emission spectrum gradually smoothening and migrating to higher energies as one approaches larger couplings \cite{CaronHuot:2006te,Hassanain:2011ce}.

\section{Concluding remarks\label{sec:conclu}}

It is clear that within 15 pages, one can only scratch the surface of a topic as extensive as theoretical heavy ion physics. In the review article at hand, we have consciously decided to not even attempt to cover all aspects of the field, but rather take a more in-depth look at a small number of recent developments that can be characterized as particularly interesting or impressive. As a guiding principle in the choice of these topics we have concentrated on first principles approaches, which oftentimes implies limited phenomenological relevance of the results, but at the same time guarantees that they are systematically improvable. Even with this bias, we have, however, only been able to cover a small fraction of relevant topics and articles, and must thus apologize for the many important works omitted. For a much more comprehensive account of the current status of the field we refer the reader to \cite{Brambilla:2014jmp}.

Beyond personal preference, the reason for our first principles bias originates from our desire to emphasize the fact that heavy ion theory is slowly but steadily reaching a stage, where a combination of first principles approaches and controlled effective descriptions is able to cover all stages of the collision process. In this article, we have tried to highlight a few of the most pertinent challenges, where we feel that significant progress has recently taken place and can be expected to continue doing so for the coming years. These include most importantly the early dynamics of the collision at realistic energies and couplings, the bulk thermodynamic properties of quark gluon plasma both at zero and finite density, as well as a reliable quantitative determination of the transport properties of the system.

\end{document}